\documentclass[12pt]{revtex4}
\usepackage{epsfig}
\usepackage{graphics}
\usepackage{subfigure}
\usepackage{amssymb,amsmath}

\begin{document}



\def\e{\begin{equation}}
\def\f{\end{equation}}
\def\_#1{{\bf #1}}
\def\o{\omega}
\def\.{\cdot}
\def\x{\times}
\def\E{\epsilon}
\def\H{\Upsilon}
\def\A{\alpha}
\def\B{\beta}
\def\O{\Xi}
\def\M{\mu}
\def\D{\nabla}
\newcommand{\ds}{\displaystyle}
\def\l#1{\label{eq:#1}}
\def\r#1{(\ref{eq:#1})}
\def\=#1{\overline{\overline #1}}
\def\##1{{\bf#1\mit}}

\title{Super-Planckian far-zone thermal emission from asymmetric hyperbolic metamaterials
}

 \author{Igor S. Nefedov}
 \affiliation{Aalto University,
School of Science and Technology \\
Department of Radio Science and Engineering\\
SMARAD Center of Excellence,
P.O. Box 13000, 00076 Aalto, Finland}
 \author{L. A. Melnikov}
 \affiliation{Yuri Gagarin State Technical University of  Saratov,
Institute of Electronics and Machinery\\
Department of Instrumentation Engineering\\
77, Politekhnicheskaya, 410054, Saratov, Russia}

 \date{\today}

 \begin{abstract}
In this paper we demonstrate that {\it asymmetric hyperbolic metamaterials (AHM)} can produce strongly directive thermal emission in far-field zone, which exceeds Planck's limit. Asymmetry is inherent in an uniaxial medium, whose optical axes are tilted with respect to medium interfaces and appears as a difference in properties of waves, propagating upward and downward with respect to the interface. It's known that a high density of states (DOS) for certain photons takes place in usual hyperbolic metamaterials, but emission of them into a smaller number in vacuum is preserved by the total internal reflection. However, the use of AHM enhance the efficiency of  coupling of the waves in AHM with the waves in free space that results in Super-Planckian far-field thermal emission in certain directions. Different plasmonic metamaterials can be used for realization of AHM. As example, thermal emission from AHM, based on graphene multilayer, is discussed.

 \end{abstract}
\pacs{44.40.+a,41.20.Jb,42.25.Bs,78.67.Wj}

\maketitle


Thermal emitters, such as a black body, typically are considered as incoherent light sources.
However, in the beginning of 2000th selective, partially coherent thermal emission, was demonstrated in far zone from three-dimensional photonic crystals \cite{Lin2000} and the grating \cite{Greffet}.
Starting from these works many structures were proposed for creation of directive thermal emitters \cite{Narayan}-\cite{Alu}. These relatively simple sources of coherent emission can find applications in photonics as alternatives to such mid-infrared sources as light-emitting diodes and quantum cascade lasers.

Since works by Kirchhoff (1859-1862) and Planck (1900) black body (BB) was considered as a perfect thermal emitter whose emissivity depends only on a temperature at a certain frequency and this limit cannot be exceeded. Rytov \cite{Rytov} (1959) and Polder and Van Hove \cite{Polder} (1971) have shown that the black-body limit for the thermal radiation heat transfer between two hot bodies, separated by nanometer gaps, can be strongly exceeded due to photon tunneling.
This effect has found applications in near-field thermophotovoltaic
systems (e.g. \cite{1,2,3}) which are often considered  as a promising tool for the field recuperation from high-temperature sources and as precise
temperature profile sensors \cite{1}. J. Pendry (1999) has shown that the dramatic enhancement holds due to excitation of coupled surface-plasmon polaritons at interfaces of two media \cite{Pendry}.
Nefedov and Simovski (2011) \cite{ThermoPRB} proposed a way to extend this effect from nanoscale to micrometer distances using media, possessing hyperbolic dispersion, i.e. hyperbolic media (HM).
Now exploitation of HM seems to be promising for enhancement of near-field thermal radiative heat transfer \cite{Biehs}-\cite{OE}.
However, according to existing concepts, the black body restriction cannot be overcome in far zone. In 2003 Lin, Moreno and Fleming reported that the BB limit was exceeded in their 3D photonic crystal emitter \cite{Lin2003}. Then Trupke et al. \cite{Trupke} criticized this work asserting that there is a violation of the second law of thermodynamics:
``There is no doubt that the density of states for certain photons and with it the photon density can be very large in a photonic crystal. The problem is to get these photons out. Emission of all photons into the smaller number of states in a vacuum would require an increase of the occupation probability, equivalent to reduced entropy. Nature avoids this violation of the second law by totally reflecting those photons, which have no corresponding states in the outside medium"Е ``What we claim here is that the photon flux thermally emitted by a body in a certain direction and in a given photon energy interval, whatever the geometry of that body, cannot exceed the photon flux thermally emitted by a black body of the same temperature, in much the same way as the absorbency of a black body, cannot further be improved to exceed unity" \cite{Trupke}. This concept regarding the far zone thermal radiation is commonly accepted up to now.

However, Kirchhoff's law was formulated in geometric optics approximation, for the condition of thermal equilibrium and for the condition of zero transverse coherence \cite{Rytov}.
We do not discuss here whether the Super-Planckian emission was really observed in \cite{Lin2003}. We refer a reader to the old Rytov's paper \cite{Rytov2} where he has shown theoretically that a sphere with diameter, compared with the wavelength, can radiate  	
several times more than BB. Another example, given in that paper, relates to thermal emission by a hot conductive wall in the rectangular waveguide. It was predicted a very strong exceed of the BB radiation near cutoff frequencies of propagating modes.
Recently Super-Planckian far-zone emission was demonstrated experimentally from a surface, covered with a transparent (non-emitting) lens, which transforms a part of spatial spectrum of evanescent waves into propagating in free spaces \cite{Fan}. Note, that the reported way of enhancing far-field thermal emission contradicts to above argumentation by Trupke because some photons, ``which have no corresponding states in the outside medium" can be emitted into free space via the lens.
In this paper we will show analytically that a very high density of states of photons can be achieved in HM and that it's possible to get these photons out without total internal reflection exploiting asymmetric hyperbolic metamaterials.

Let us discuss the difference between non-magnetic homogeneous isotropic medium (IM), having dielectric  constant $\epsilon$ and uni-axial homogeneous anisotropic (hyperbolic) medium HM with dielectric tensor $\{\{\epsilon_t,0,0\},\{0,\epsilon_t,0\},\{0,0,\epsilon_z\}\}$. For both media DOS is $dN=(2\pi)^{-3}k^2 dk d\Omega$, where only one (extraordinary) polarization is considered, $k$ is the wavenumber, and $d\Omega=\sin{\theta}d\theta d\Phi$ is the elementary solid angle. Dispersion equation for IM and HM  are correspondingly
$\o_{IM}(k)=ck\epsilon,\,\,\o_{HM}(k)=ck \sqrt{\frac{1}{2}[\epsilon_t^{-1}+\epsilon_z^{-1}+(\epsilon_t^{-1}-\epsilon_z^{-1})\cos{\theta}]}$,
\cite{Felsen}. The density of thermal radiation in IM and HM is $\rho_{IM,HM}=dN\Theta(\omega,T)$, where $\Theta(\omega,T)=\hbar\o_{IM,HM}(k)/[\exp(\frac{\hbar\o_{IM,HM}(k)}{T})-1]$ is the average energy of Planck's oscillator. Here $T$ is the absolute temperature in energy units. The thermal radiation is not isotropic in anisotropic medium \cite{Mer,Chug}. To compare the results for IM and HM let us integrate these expressions on $k$ from $0$ to $\infty \,(\hbar=1,c=1)$. The ratio of integrals is ${2 \sqrt{2}} {\epsilon^{-1}\left[\cos{2\theta} \left({\epsilon_t}^{-1}-{\epsilon_z}^{-1}\right)+{\epsilon_t}^{-1}+{\epsilon_z}^{-1}\right]^{-3/2}}.$ Let us suppose that $\epsilon=\epsilon_t$. Then for anisotropic media with $\epsilon_{t,z}>0$ this ratio is less than 1 for near all values of $\theta$.
For hyperbolic media with $\epsilon_t \epsilon_z <0$ thermal radiation exceeds the radiation in IM for all values of $\theta$. The reason is  high DOS in HM for the same frequencies in comparison with DOS in IM.

Then let us suppose that IM and HM are separated by the interface plane. Thus the waves can propagate in both media, depending on the dielectric constants and interface orientation according to the optical axis of HM. To calculate the characteristics of thermal radiation the eigenmodes of the whole structure is necessary to find, together with dispersion relation for frequency and density of modes \cite{Kroll}. Let us consider the closed cavity with ideally reflecting walls in the form of cuboid $l_x \times l_y\times l_z$, in which IM occupies the volume $-l_z/2<z<0$ while HM fill the rest of the cuboid $ 0<z<l_z/2$. The optical axis of HM is tilted according to the interface plane by the angle $\phi$. The field of the structure can be presented as plane waves with wave vector $\vec k=(k_x,k_y,k_z)$ in IM-part and plane waves with wave vector $\vec k_{HM}=(k_x,k_y,k_z^{HM})$ in HM-part.  Due to interface at $ z=0$ these waves become coupled forming eigenmodes of whole structure.  Each mode "lives" both in IM and HM, and quanta of field are spread among the media depending on interface orientation and $\vec k$, which parametrises the eigenmodes. Providing the field oscillators are thermally excited, the equilibrium state is achieved with Planck's average energy of oscillator,  however the spatial distribution and energy can be other than in BB radiation. To find DOS of the structure the dispersion equations (for extraordinary wave) should be used: $\o_{IM}^2=\o_{HM}^2 \rightarrow k_{z,\,1,2}^{HM}=\sec{2 \phi} (\pm\sqrt{k_x^2 + (k_x^2 + 2 k_y^2 + k_z^2) \cos{2 \phi}} - k_x \sin{2 \phi})$ for $\epsilon_t=1,\,\epsilon_z=-1$. The accumulated phase during propagation from $-l_z/2$ to $ l_z/2$ must be equal to $2\pi n_z$ (periodic boundary conditions): $(l_z/2) (k_z+k_{z,1}^{HM}(k_x,k_y,k_z))= 2\pi n_z,\,2\pi n_z=(l_z/2) (k_z+k_{z,2}^{HM}(k_x,k_y,k_z))$ for downward and upward waves. When $k_z>>k_x,k_y$ the solution of this equations is $(2\pi)^{-1}k_z\approx 2 n_z(\sqrt{\cos{2\phi}}-\cos{2\phi})/(1 - \cos{2 \phi})$ which can be used for calculating of DOS: $dN=dn_x dn_y dn_z=(2\pi)^{-3}dk_x dk_y dk_z (1 - \cos{2 \phi})/(\sqrt{\cos{2\phi}}-\cos{2\phi}),\,l_x l_y l_z=1$. It means the DOS goes to infinity when $\phi \rightarrow \pi/4$. It is easy to check that for anisotropic medium with $\epsilon_t>1, \epsilon_z>1 $ DOS becomes less than in vacuum due to total internal reflection at the interface because we use $\vec k$ in vacuum as the mode parameter, rather than  $\vec k_{HM}$. Thus combining IM and AHM we can obtain the medium with higher DOS corresponding propagating modes than in vacuum only. Field energy distribution among IM and AHM parts  depends on the reflection at the interface plane.
\begin{figure}[!h] 
\subfigure[]{\includegraphics[width=0.46\linewidth]{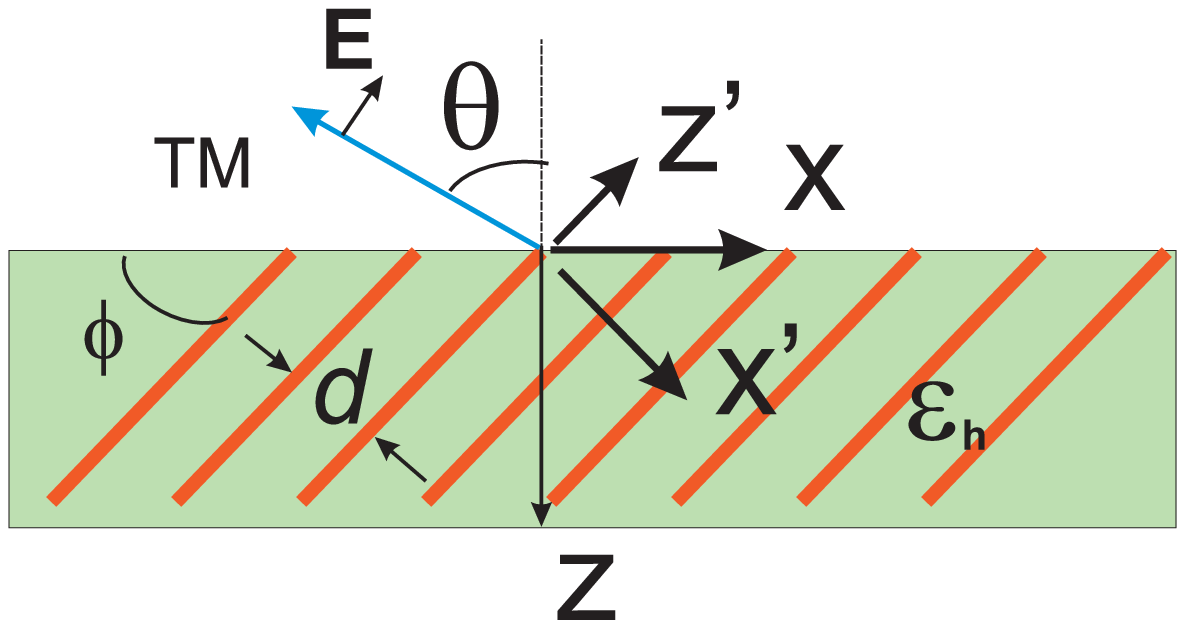}}
\subfigure[]{\includegraphics[width=0.46\linewidth]{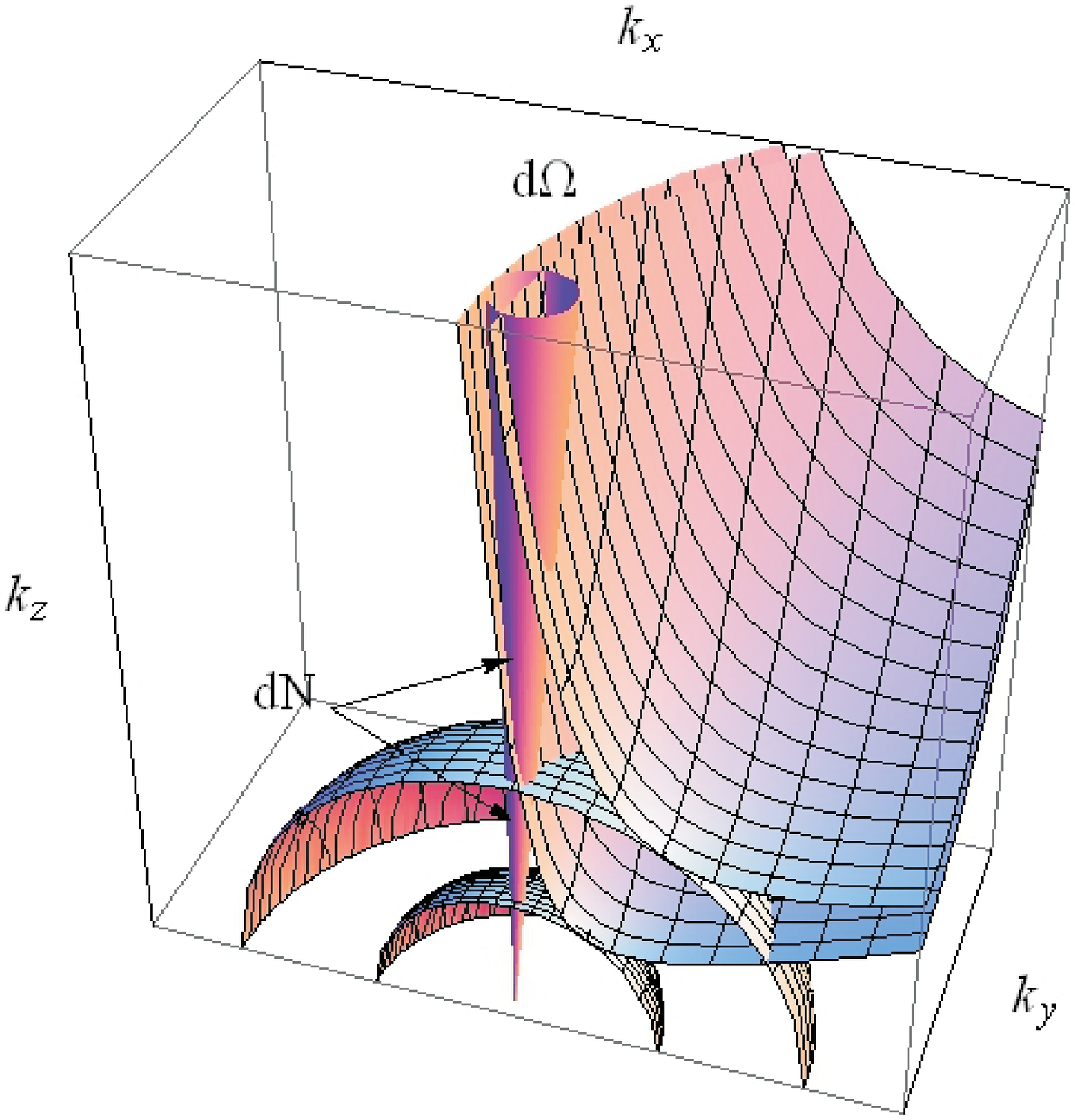}}
\caption{(a) Schematic view of a hyperbolic metamaterial
with a tilted optical axis. (b) Elliptic and hyperbolic dispersion surfaces in 3D space, corresponding to the same frequency. Parameters of AHM are the following: $\E_t=1$, $\E_z=-1$.  }
 \label{tt}
 \end{figure}

In asymmetric hyperbolic metamaterials, i.e. metamaterials with anisotropy axes, tilted with respect to interfaces, the reflection can be made small enough to manifest DOS increase. Such metamaterials are characterized by an asymmetry in properties of waves, propagating upward and downward with respect to AHM interfaces under a fixed transverse component of the wave vector, and exhibit unique absorbing properties (see \cite{Has}-\cite{graphene}).
Fig.~\ref{tt}b shows spherical isofrequency surfaces (for air) and hyperbolic ones, calculated for the same frequency. The density of states $dN$ is proportional to a volume, enclosed in a solid angle $d\Omega$ between two isofrequencies. Projection of this volume onto the $(k_xk_y)$-plane belongs to the area of propagating waves. It means that the waves with very large modules of the wave vector can leave AHM without the total internal reflection and, in other words, high DOS photons in AHM are coupled  with photons in free space.
Regarding a usual reflection, as we will show below, it is quite moderate and does not suppress the far-field SP emission.

Schematic view of AHM is shown in Fig.~\ref{tt}a.
We assume that the medium is semi-infinite in the $-z$ - direction and neglect its possible spatial dispersion.
The permittivity tensor in the $XOZ$ coordinate system, associated with the medium interface, can be expressed through rotation transformation and
the cartesian components of $\=\E$ read:
\e\l{e5}\begin{array}{l}
\E_{xz}=\E_{zx}=(\E_{z}-\E_t)\cos{\phi}\sin{\phi}\\
\E_{xx}=\E_{z}\sin^2{\phi}+\E_t\cos^2{\phi}\\
\E_{zz}=\E_{z}\cos^2{\phi}+\E_t\sin^2{\phi}.\end{array}\f

Since conventional formulas expressing thermal emission (see, for example \cite{Vol1}-\cite{volokitin}) are not applicable for media, characterized by non-diagonal tensors of the permittivity, we start as in \cite{Polder} from nonhomogeneous Maxwell's equations with random electric current density sources $\_j$ and
apply the fluctuation-dissipation theorem
\cite{Landau} for the ensemble-averaged bulk current density:
\e\l{fdt}
\begin{array}{lc}
\langle j_m(\_r,\omega)j^*_n(\_r',\omega')\rangle=  \\
 \frac{4}{\pi}\omega\E_0\E_{mn}''(\omega)\delta(\_r-\_r')\delta(\omega-\omega')\Theta(\omega,T).
\end{array}\f
In Eq.~\r{fdt}  $\E_{mn}''\equiv {\rm Im}(\E_{mn})$, $j_m$ and $j_n$
($m,\,n=1,\,2,$ or 3) are $x,\ y$ or $z$ component of $\_j$,
$\delta(x)$ is the Dirac
delta function, $\E_0$ is the permittivity of vacuum.
Omitting time dependence $e^{-i\omega t}$, an elementary bulk current source can be written as
\e\l{f1}
\_j(x,z)=\_j_0(x',z')\delta(\bar{x})\delta(\bar{z}),\f
where $\bar{x}=x-x',\;\bar{z}=z-z'$.

For any uniaxial crystal Maxwell equations are split into two sub-systems describing the ordinary and extraordinary waves.
A hyperbolic electric medium supports propagation of TM (p-polarized) waves only for a whole spectrum of transversal wave vectors, which are extraordinary ones.
We restrict our consideration by the TM-waves propagating in the $(xz)$-plane ($\partial/\partial y=0$).
In this case the electric field vector lies in plane of the anisotropy axis which is orthogonal to the interface, and both reflected and transmitted waves keep the TM-polarization.

Let us use the spectral representation of current and nonzero field components of the TM-polarized wave in bulk AHM:
\e\l{EH}
\left[\begin{array}{c}
\_j(x,z) \\
E_{x}(x,z)\\
H_y(x,z)
\end{array}\right]=
\int\int_{-\infty}^{\infty}\left[\begin{array}{c}
\_j_0(x',z')\\
e_{x}(\A,\B)\\
h_{y}(\alpha,\B)
\end{array}\right]
\frac{e^{i[\alpha (\bar{x})+\B (\bar{z})]}}{(2\pi)^2}\,d\alpha d\B.
\f
Here
we included exponent $e^{-i(\alpha x'+\B z')}$ into unknown Fourier transforms
$e_{x}(\A,\B),\;h_{y}(\alpha,\B)$ for convenience taking into account Fourier transform of the delta-functions in \r{f1}.
In the Fourier space
fields, created at point $(x,z)$ by the source located at $(x',z')$ can be expressed as follows:
\e\l{J}\begin{array}{c}
e_x(\alpha,\B)=\frac{i}{\Delta}[-j_{0x}(x',z')A_{22}+
j_{0z}(x',z')A_{12}]\\
h_y(\alpha,\B)=\frac{i}{\Delta}[-j_{0x}(x',z')A_{21}+
j_{0z}(x',z')A_{11}]
\end{array}\f
where $\rho=120\pi$ is the wave impedance of vacuum,
\e\l{del}\Delta=A_{11}A_{22}-A_{12}A_{21},\f
\e\l{A}\begin{array}{cc}
A_{11}=\frac{k_0}{\eta}\left(\E_{xx}+\frac{\B}{\alpha}\E_{xz}\right) &
A_{12}=-\left(\B+\frac{k_0^2}{\alpha}\E_{xz}\right) \\
A_{21}=\frac{k_0}{\eta}\left(\E_{xz}+\frac{\B}{\alpha}\E_{zz}\right) &
A_{22}=\alpha-\frac{k_0^2}{\alpha}\E_{zz}.
\end{array}\f
Note that condition $\Delta(\alpha)=0$ is actually the dispersion equation for the TM waves in unbounded medium.

Let us substitute \r{J} into Eq.~\r{EH} and
evaluate integral over $\beta$ using residuum method.
Poles of integrands correspond to zeros of equation
\e\l{D}
\Delta(\B_m)=0\f
which read as \cite{Felsen,Has}
\e\l{kz}
\B_{1,2}=
k_z^{(1,2)}=\frac{\A\E_{xz}\pm\sqrt{(\E_{xz}^2-\E_{xx}\E_{zz})(\A^2-k_0^2\E_{zz})}}{\E_{zz}}\f
(Some interesting features of these solutions for hyperbolic media are described in \cite{Felsen}-\cite{ScRep}).
We close integration contour in the lower half-space in the complex plane,
choosing such a root of \r{D} that $\Im(k_z^{(i)})<0$ corresponds to the wave whose amplitude attenuate propagating in the positive direction of the $z$-axis at $z<0$.

Applying the residuum method we come to the following expressions for field components at the point $(x,z)$, excited by the point-like current, located at the point $(x',z')$:
\e\l{EH1}
\left[\begin{array}{c}
E_{x}\\ 
H_y     
\end{array}\right]=
\int_{-\infty}^{\infty}\left[\begin{array}{c}
\tilde e_{x}(\A,z)\\
\tilde h_{y}(\A,z)
\end{array}\right]e^{\A(\bar{x})}
\frac{e^{-ik_z^{(m)}|\bar{z}|}}{2\pi}\,d\A
\f
\e\l{J1}\begin{array}{c}
\tilde e_x(\A,z)=\frac{1}{\Delta'}[-j_{0x}(x',z')\tilde A_{22}(\A)+j_{0z}(x',z')\tilde A_{12}(\A)]\\
\tilde h_y(\A,z)=\frac{1}{\Delta'}[-j_{0x}(x',z')\tilde A_{21}(\A)+j_{0z}(x',z')\tilde A_{11}(\A)]
\end{array}\f
where where Im$(k_z^{(m)})\leq 0$ and expressions for $\tilde A_{ij}$ are obtained from \r{A} substituting $k_z(\A)$ instead $\B$.
Here superscript $m$ is omitted in notation for $k_z$ and
\e\l{d1}
\Delta'(k_z^{(m)})=\frac{d}{d\B}\Delta(\B)_{|\B=k_z^{(i)}}=
\frac{2k_0}{\eta}\left(\E_{xz}+\frac{k_z}{\A}\E_{zz}\right)
\f
Thus, formulas \r{EH1},\r{J1} give us actually expressions for corresponding components of dyadic Green's functions.

Then we compose from field components the quadratic form
$E_x(x,z|x',z')H_y^*(x,z|x'',z'')$, corresponding to the Poynting vector,
and integrate it over all distributed sources, located in points $(x',z')$ and $(x'',z'')$.
Evaluating integration and using \r{fdt} we obtain expression for
the ensemble-averaged Poynting vector, incident onto the interface $z=0$:
\e\l{Sz}
\langle S_z(x,\omega)\rangle=\int_{-\infty}^{\infty}\,\langle S_z(k_x,\omega)\rangle\,dk_x\f
where
\e\l{Poyn}\begin{array}{l}
\langle S_z(k_x,\omega)\rangle=\frac{k_0}{\pi\eta}\frac{\Im(k_z)}{|\Delta'|^2}\left[
\E_{xx}''\tilde A_{22}\tilde A_{21}^*+  \right. \\
+\left. \E_{zz}''\tilde A_{12}\tilde A_{11}^*-
\E_{xz}''(\tilde A_{22}\tilde A_{11}^*+\tilde A_{12}\tilde A_{21}^*)\right].
\end{array}\f

Transmission coefficient for the Poynting vector reads \cite{ThermoPRB}
\e\l{T}
t_s=|t|^2Z_t^*/Z_0^*, \f
where $Z_0$ and $Z_t$ are transverse wave impedances for vacuum and  the asymmetric hyperbolic medium, respectively. Expressions for them are given in \cite{Has,ScRep}.
Note, that $Z_t$, corresponding to $k_z^{(1,2)}$, differ only in sign.
Thus, the spectral density of thermal emission flux, radiated at angle $\theta$ into a plane angle
$d\theta$ can be written as
\e\l{P}
q(\omega,\theta)=\langle S_z(k_x,\omega)\rangle t_s\cos{\theta}+c.c.\f


As example of AHM we consider a graphene multilayer (GM) which is the structure consisted of periodically arranged graphene sheets embedded into a host matrix with the relative permittivity $\E_h$ and tilted with respect to interface.
For the relative transverse tensor component $\E_t$ we used the homogenization model, described in \cite{graphene}. Graphene conductivity was calculated by the Kubo formula \cite{Hanson}, and $\E_z=\E_h$. Under certain conditions graphene multilayers exhibit properties of hyperbolic media \cite{graphene,Iorsh}.

\begin{figure}[!h] 
\subfigure[]{\includegraphics[width=0.45\linewidth]{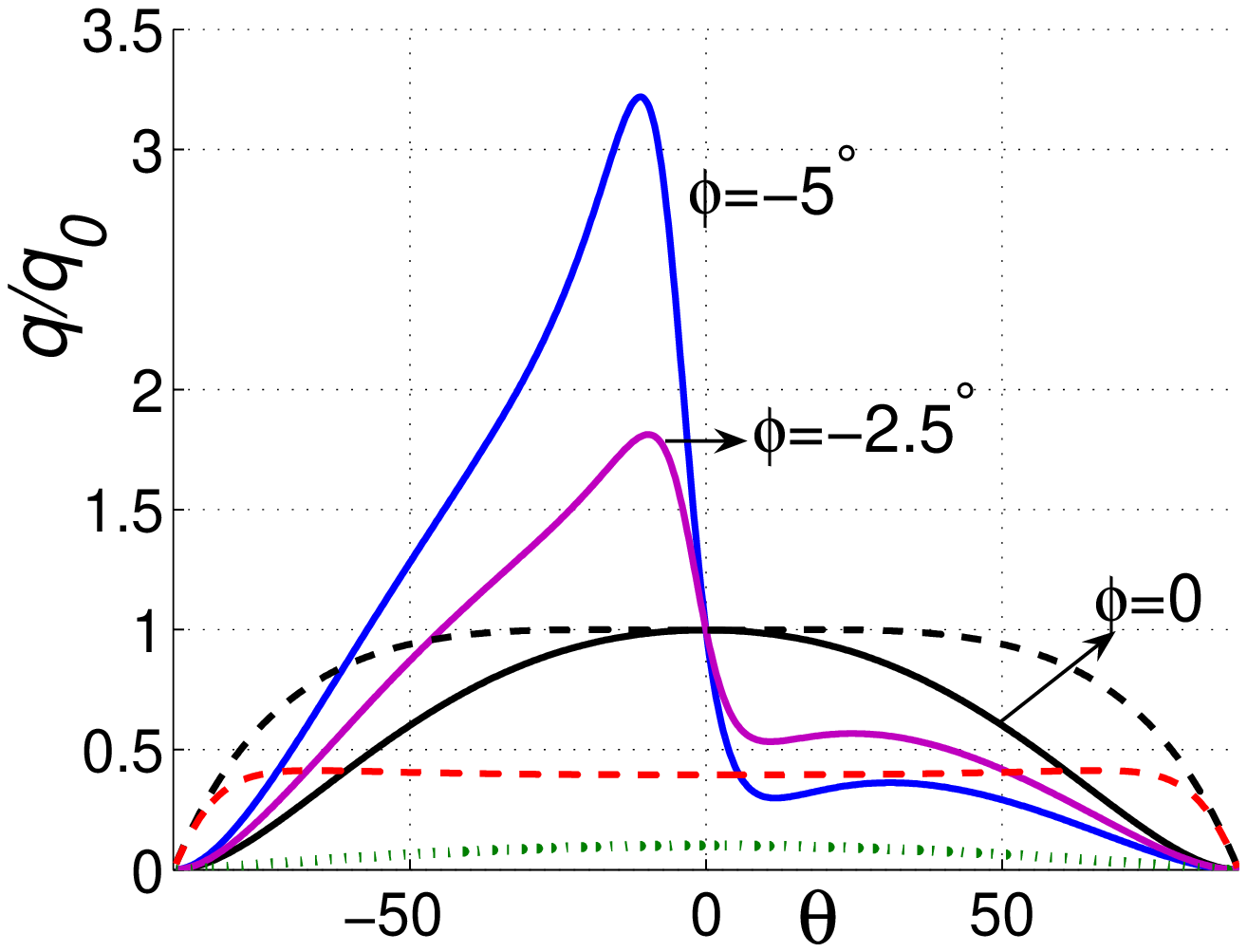}}
\subfigure[]{\includegraphics[width=0.47\linewidth]{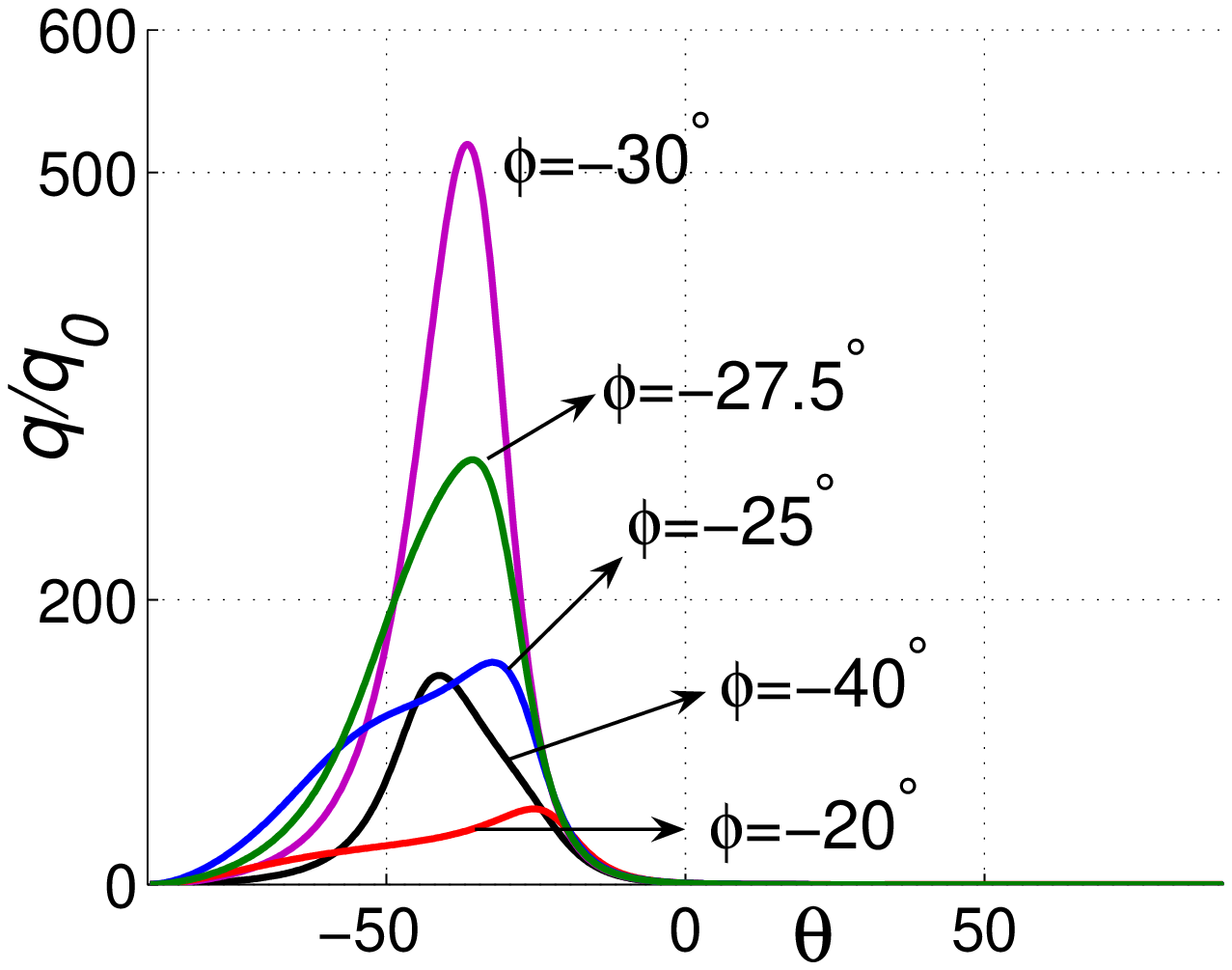}}
\caption{(Color online) Normalized emissivity, calculated for different tilt angles $\phi$.
Dashed curves show modulus of transmission for the Poynting vector. Dotted green curve shows normalized emissivity of the TE-polarized wave.}
 \label{sh}
 \end{figure}

Figs.~\ref{sh}a,b illustrate the spectral density of the thermal emission flux from graphene multilayers with different tilt angles $\phi$ into a plane angle.
Thermal emission is normalized on the same characteristic of the black body $q_{\rm BB}(\omega,\theta)$. Parameters of GM are the following: $d=10$\,nm; $\E_h=1.2+i0.01$;
the relaxation time of carriers is taken to be $10^{-13}$\,s, the chemical potential of graphene $\mu_c=0.15$\,eV, that can be provided by applying a voltage, and the wavelength $\lambda=8\,\mu$m.
Under such parameters $\E_{\perp}=-1.3617 + i0.7622$.
Fig.~\ref{sh}a shows emission, produced by GM with small tilt angles and with vertically standing sheets (red curve), so the permittivity tensor is a diagonal one or having small non-diagonal components. Transmission coefficient for the Poynting vector, defined by formula \r{T}, also is shown.

\begin{figure} 
\subfigure[]{\includegraphics[width=0.50\linewidth]{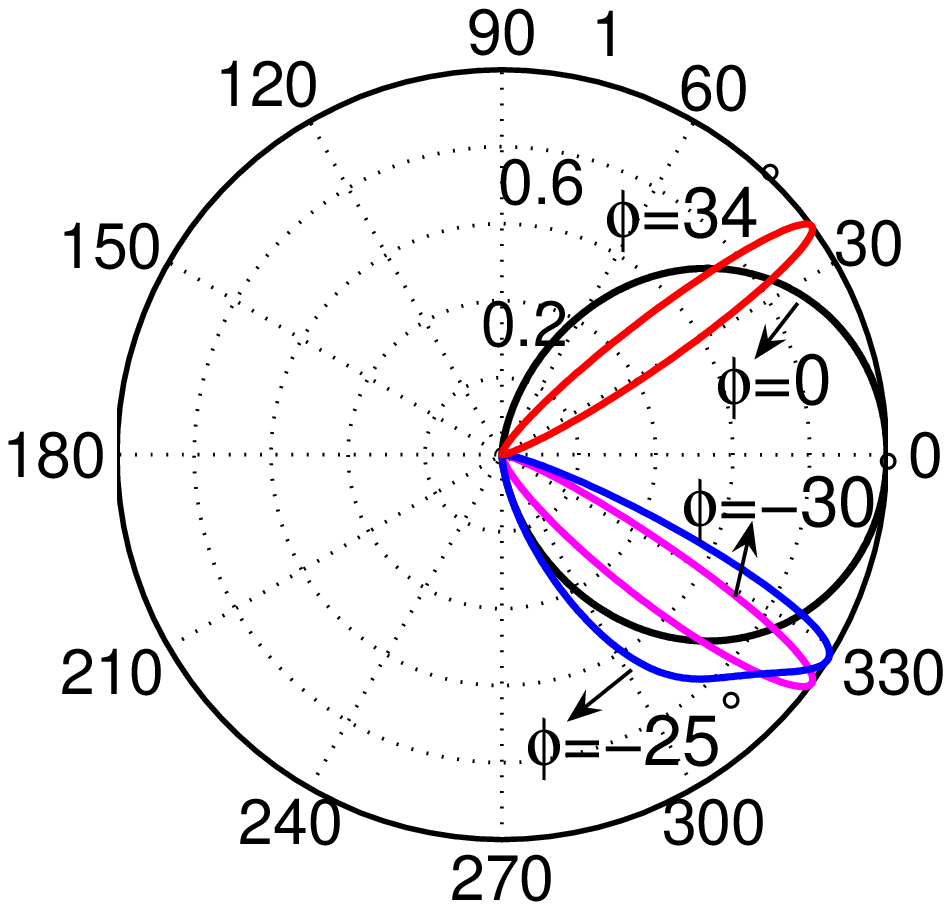}}
 \subfigure[]{\includegraphics[width=0.47\linewidth]{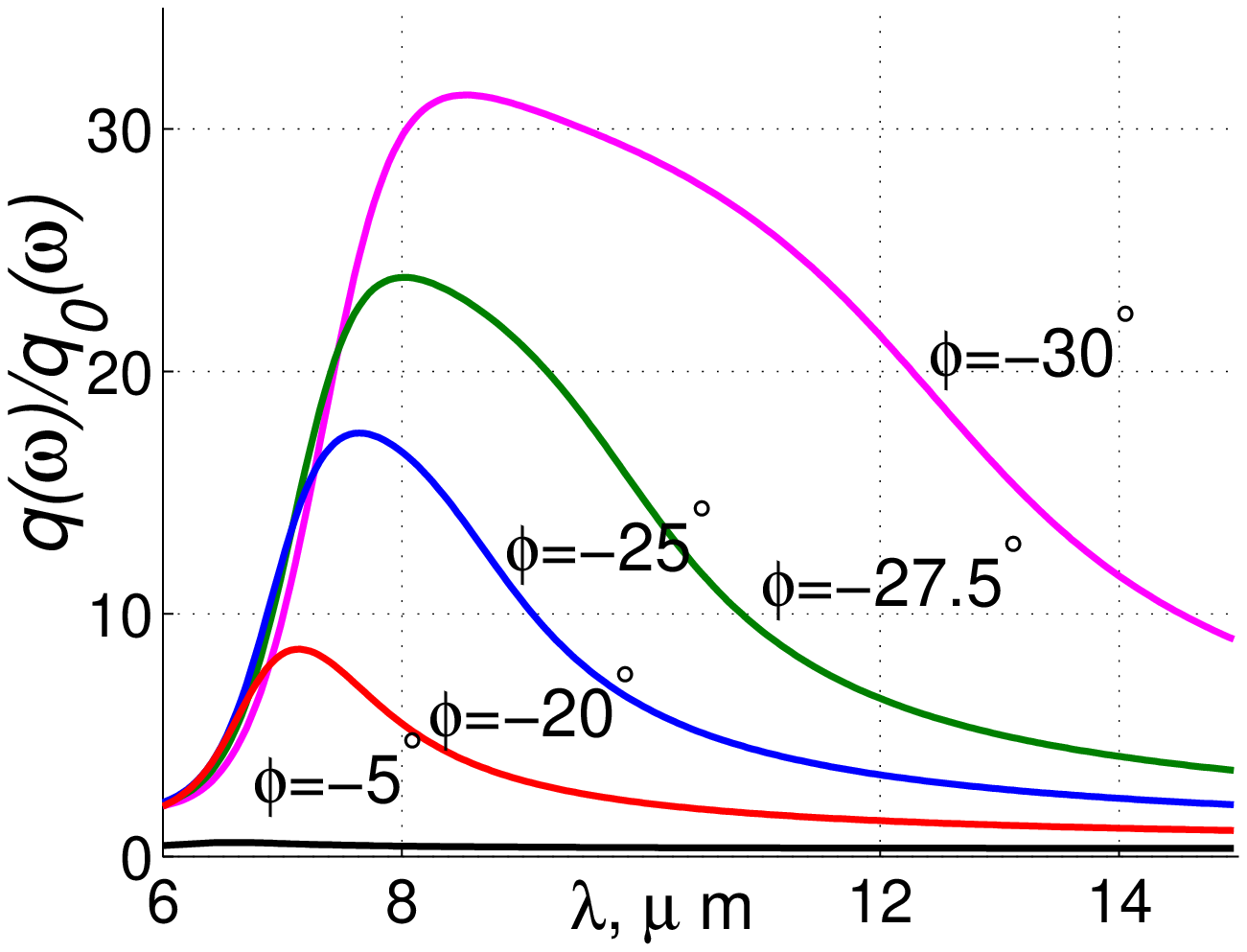}}
\caption{(Color online) (a) Wavelength dependence of the normalized emissivity, calculated for the different tilt angles. (b) Radiation pattern in polar coordinate, calculated for the same GM parameters and different tilt angles.}
 \label{total}
 \end{figure}

For the tilt angles $\phi=0,\,2.5^{\circ}$ and $5^{\circ}$ transmission coefficients are very close and shown by black dashed curve. For $\phi=90^{\circ}$ transmission is much less (see red dashed curve).
Super-Planckian radiation appears already at small tilt angle $2.5^{\circ}$ and its maximum increases with $\phi$.
Thermal emission from GM at larger $\phi$ is much higher, see  Fig.~\ref{sh}b. There is an optimum which depends on  parameters of GM. Emissivity of the TE-wave also is shown in Fig.~\ref{sh}a. For its calculation known formulas from \cite{Polder} can be used.
For illustration, the directivity of thermal emission in polar coordinates is shown in Fig.~\ref{total}a.



Mathematical explanation of SP radiation follows from formula \r{Poyn}. Angular maxima of emission correspond to minima of $|\Delta'(\theta)|$ which are the close to zero the material losses are lower.
This situation takes place in any media. Let us consider an isotropic medium with the scalar permittivity $\E$.
In a plane wave expansion form the field components, excited by a point-like source in the medium, are inversely proportional to $\sqrt{k_0^2\E-k_x^2}$ \cite{Felsen}. So, the $k_x$-Fourier component goes to infinity if
$(\E'')\rightarrow 0$ and $|k_x|\rightarrow k_0\sqrt{\E}$.
However waves, belonging to a spatial spectrum area $k_x\approx k_0\sqrt{\E}$, undergo total internal reflection at an interface with vacuum.
One can show that similar situation takes place in anisotropic media if anisotropy axes either parallel or orthogonal to media interfaces. In contrast, in AHM the minimum of $|\Delta'(k_x)|$ falls to the area of propagating in vacuum waves, $|k_x|<k_0$ and, similarly to the isotropic case,
$\langle S_z(k_x,\omega)\rangle\rightarrow\infty$ at the minimum of $|\Delta'(k_x)|$ if Im$(\E_z)\rightarrow 0$, Im$(\E_t)\rightarrow 0$.


Fig.~\ref{total}b shows thermal emission of TM-polarized waves into the plane angle in the $(xz)$-plane, integrated over the emission angle $\theta$, $-90^{\circ}\leq\theta\leq 90^{\circ}$. Thus, we obtained emission into a full plane angle, exceeding black body spectrum by factor 30.

Finally we will discuss whether Kirchhoff's law is violated.
Now it is well-known
that a plasmonic resonant particle can absorb more than the light incident on it \cite{Bohren,Paul}.
So strong absorption is associated with excitation of surface plasmons or surface phonons \cite{Bohren} and it is inversely proportional to $\E''$ at a resonant frequency.
(This is taking place in the case of thermal emission from AHM).
One can note that the reverse side of ``absorption more than 100\%" is the super-Planckian thermal emission, produced by the same particle in the far-field zone. For this case the conventional Kirchhoff's law was generalized introducing the effective absorption cross-section \cite{Rytov}.
After this generalization Kirchhoff's law is not violated despite thermal emission from a unit square exceeds the black body limit.
Similar effect does not present in conventional bulk media due to impossibility to excite slow surface plasmonic polaritons by fast waves incoming from free space. However, bulk plasmons can be excited
in asymmetric hyperbolic media by external radiation \cite{Has}-\cite{graphene}.
Effect of super-Planckian radiation from AHM does not violate thermodynamics laws, similarly as SP radiation described in \cite{Rytov2,Fan}. Enhanced emission requires increased thermal flow supply to a radiating body in order to maintain a constant temperature.

Concluding we note that our results are based on the analytical approach which exploits a quite simple scheme, described by Polder \& Van Hove, and can be easily reproduced.


\end{document}